\documentclass[10pt,sigconf,letterpaper]{acmart}
\usepackage{listings}
\usepackage{pifont}
\usepackage{adjustbox}
\usepackage{caption}
\usepackage{subcaption}
\usepackage{xcolor}
\usepackage{multirow}

\newcommand{\xmark}{\ding{55}}

\AtBeginDocument{%
  \providecommand\BibTeX{{%
    \normalfont B\kern-0.5em{\scshape i\kern-0.25em b}\kern-0.8em\TeX}}}

\setcopyright{acmcopyright}
\copyrightyear{2022}
\acmYear{2022}
\acmDOI{XXXXXXX.XXXXXXX}

\acmConference[DataEconomy@CoNEXT'22]{ACM Data Economy Workshop co-located with the 2022 International Conference on emerging Networking EXperiments and Technologies}{December 6--9, 2022}{Rome, Italy}
  
\acmPrice{15.00}
\acmISBN{978-1-4503-XXXX-X/18/06}

\begin{document}

\settopmatter{printacmref=false}

\title{Towards Employing Recommender Systems for Supporting Data and Algorithm Sharing}

\author{Peter Müllner}
\affiliation{%
  \institution{Know-Center Gmbh}
  \city{Graz}
  \country{Austria}
}
\email{pmuellner@know-center.at}

\author{Stefan Schmerda}
\affiliation{%
  \institution{Know-Center Gmbh}
  \city{Graz}
  \country{Austria}
}
\email{sschmerda@know-center.at}

\author{Dieter Theiler}
\affiliation{%
  \institution{Know-Center Gmbh}
  \city{Graz}
  \country{Austria}
}
\email{dtheiler@know-center.at}

\author{Stefanie Lindstaedt}
\affiliation{%
  \institution{Know-Center Gmbh \& TU Graz}
  \city{Graz}
  \country{Austria}
}
\email{slind@know-center.at}

\author{Dominik Kowald}
\affiliation{%
  \institution{Know-Center Gmbh \& TU Graz}
  \city{Graz}
  \country{Austria}
}
\email{dkowald@know-center.at}

\renewcommand{\shortauthors}{Müllner, Schmerda, Theiler, Lindstaedt, and Kowald}

\begin{abstract}
Data and algorithm sharing is an imperative part of data- and AI-driven economies. The efficient sharing of data and algorithms relies on the active interplay between users, data providers, and algorithm providers. Although recommender systems are known to effectively interconnect users and items in e-commerce settings, there is a lack of research on the applicability of recommender systems for data and algorithm sharing. To fill this gap, we identify six recommendation scenarios for supporting data and algorithm sharing, where four of these scenarios substantially differ from the traditional recommendation scenarios in e-commerce applications. We evaluate these recommendation scenarios using a novel dataset based on interaction data of the OpenML data and algorithm sharing platform, which we also provide for the scientific community. Specifically, we investigate three types of recommendation approaches, namely \mbox{popularity-,} collaboration-, and content-based recommendations. We find that collaboration-based recommendations provide the most accurate recommendations in all scenarios. Plus, the recommendation accuracy strongly depends on the specific scenario, e.g., algorithm recommendations for users are a more difficult problem than algorithm recommendations for datasets. Finally, the content-based approach generates the least popularity-biased recommendations that cover the most datasets and algorithms.
\end{abstract}






\maketitle

\section{Introduction}

\begin{figure}
     \centering
     \includegraphics[width=\linewidth]{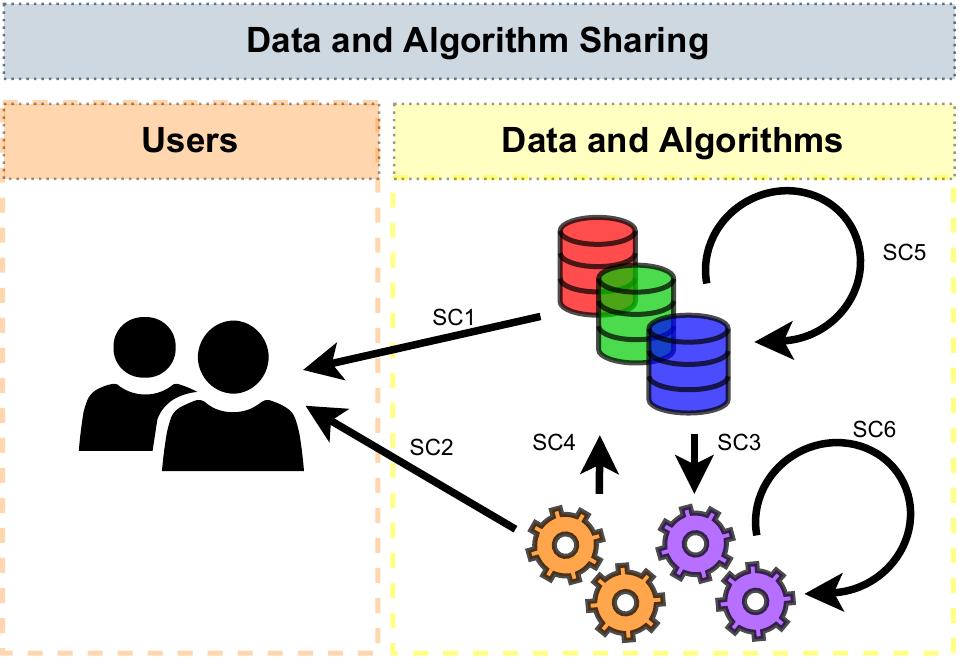}
     \caption{Recommendation scenarios SC1-SC6 that can support data and algorithm sharing. In addition to the traditional item-to-user scenarios SC1 and SC2, also item-to-item scenarios SC3-SC6 can occur.\vspace{-3mm}}
    \label{fig:scenarios}
\end{figure}

Sharing data and algorithms is one important cornerstone in today's data- and AI-driven economy.
To enable data and algorithm sharing, interconnecting three key-players is essential: data providers, algorithm providers, and users.
\emph{Data Providers} grant access to their data collections.
\emph{Algorithm Providers} allow applying their algorithms to a given piece of data.
\emph{Users} apply algorithms to data and, this way, connect data and algorithms.
In general, data and algorithm providers may share their resources due to various reasons, e.g., to monetize the data or the algorithm, or to make them available for the research community.
The powerful strength of data and algorithm sharing lies in the exploitation of shared resources, e.g., data shared by a data provider.
For example, it might be advantageous for companies to gain access to the best-suited data to enhance their AI  pipeline.
However, selecting the best-suited dataset is hard, which stems from the fact that the number of available datasets, publicly available over the Web or stored in private databases, has increased rapidly over the last decade~\cite{stahl2016classification,liang2018survey, jess2015industrial,fernandez2020data}. 

Although the deployment of recommender systems for e-commerce, e.g., Amazon or Zalando, is a natural decision to address this choice overload, not much research is available on the applicability of recommender systems for data and algorithm sharing (see Section~\ref{s:relwork}).
This is especially true for beyond-accuracy objectives of recommender systems, such as popularity bias, which is currently an important topic in the research community.
Recommender systems exhibiting popularity bias tend to exclude many datasets and algorithms from their recommendations and recommend popular items substantially more often than non-popular items~\cite{elahi2021investigating,kowald2021support,kowald2020unfairness}. 

To study to what extent recommender systems can support data and algorithm sharing, we identify six recommendation scenarios (see Figure~\ref{fig:scenarios}).
In these scenarios, we evaluate three recommendation methods, i.e., Most Popular, Collaborative Filtering, and Content-based Filtering, with respect to recommendation accuracy and popularity bias.
The three main-contributions of this paper are as follows:
\begin{enumerate}
    \item We discuss six recommendation scenarios and outline how recommender systems can be applied to support data and algorithm sharing (see Section~\ref{s:scenarios}).
    \item We create and publish a novel dataset based on the OpenML platform, which allows studying recommender systems for data and algorithm sharing (see Section~\ref{s:method}).
    \item We show that Collaborate Filtering yields the most accurate recommendations and Content-based Filtering can generate recommendations that cover the most datasets and algorithms (see Section~\ref{s:results}).
\end{enumerate}

\section{Related Work} \label{s:relwork}
Recommender systems for data and algorithms are of growing interest to both academia and industry in the field of data and AI-driven economies~\cite{patra2020content,fernandez2020data,jess2015industrial}.

For example, Patra et al.~\cite{patra2020content} utilize Content-based Filtering for dataset recommendations in the genetics domain.
Also, Jess et al.~\cite{jess2015industrial} design a recommender system for artificial data to help human decision-making in the industrial domain.
The task of algorithm recommendations has been partially approached by Automated Machine Learning, which aims to automatically select an appropriate machine learning pipeline (including algorithms) for a given dataset and problem~\cite{he2021automl}.
For example, Zschech et al.~\cite{zschech2019towards} recommend a data mining pipeline for a given problem.
Vainshtein et al.~\cite{vainshtein2018hybrid} and Song et al.~\cite{song2012automatic} exploit metadata and structural properties of datasets to recommend classification algorithms.


Numerous works exist that evaluate recommender systems for popularity bias, i.e., their inclination to recommend popular items~\cite{elahi2021investigating,zhu2021popularity,mansoury2020feedback}.
For example, Mansoury et al.~\cite{mansoury2020feedback} show that recommender systems can seriously exacerbate existing biases, such as popularity bias.
Also, Zhu et al.~\cite{zhu2021popularity} simulate a recommender system to monitor the evolution of popularity bias.
Within this dynamic setting, the authors studied factors that drive popularity bias.


Data Market Austria (DMA)\footnote{\url{https://www.datamarket.at/}} is an example of a data- and AI-driven economy, in which a recommender system is employed to connect users, data, and algorithms~\cite{kowald2019using}. 
However, the authors raise concerns regarding the dataset used in their study with respect to valid connections between users, datasets, and algorithms, and do not consider content-based recommendations.
Plus, our work includes a beyond-accuracy evaluation study with respect to popularity bias.

\section{Recommendation Scenarios} \label{s:scenarios}
Recommender systems rely on (i) profile data for model training and (ii) ground truth data for model evaluation.
In a traditional item-to-user recommendation scenario (SC1 and SC2), \emph{profile data} refers to the user profile that represents a user's item preferences.
\emph{Ground truth data} represents the user's item preferences the recommender system aims to predict.
Typically, the direct interactions between users and items (e.g., a user's utilization of a certain dataset) are used as the users' item preferences.
However, for the remaining item-to-item recommendation scenarios (SC3-SC6) there is no direct item-to-item interaction data that can be used to generate recommendations, e.g., dataset to algorithm recommendations (see Table~\ref{tab:differences}).

Thus, in the following, we detail our six recommendation scenarios that can occur in data and algorithm sharing (see Figure~\ref{fig:scenarios}) and give examples how recommender systems can cope with the lack of direct interactions for item-to-item recommendation scenarios:

\textbf{SC1: Datasets to Users.}
In SC1, recommendations help users (e.g., researchers) to identify datasets that are deemed to be relevant.
As Figure~\ref{fig:scenarios} illustrates, there exists a direct interaction between users and datasets (e.g., a user uses a dataset to train an algorithm). 
Thus, the recommender system can leverage these interactions to generate recommendations.

\textbf{SC2: Algorithms to Users.} 
In SC2, recommendations help users (e.g., researchers) to identify algorithms that are deemed to be relevant.
As in SC1, also in SC2, the recommender system can leverage the direct interactions between users and algorithms to generate recommendations. 

\begin{table}[!t]
    \centering
    \caption{Profile data is used to generate recommendations, which are then evaluated against the ground truth data. 
    In the item-to-user scenarios SC1 and SC2, profile and ground truth data are available via direct user-to-item interactions (e.g., user utilizes a dataset).
    However, for our remaining scenarios, i.e., SC3-SC6, profile and/or ground truth data is available only via indirect item-to-item interactions, or unavailable (\xmark), and needs to be constructed.}
    \begin{small}
    \begin{tabular}{l c c}
    \toprule
    & Profile Data & Ground Truth \\ \midrule
    SC1: Datasets to Users & direct & direct \\
    SC2: Alsorithms to Users & direct & direct \\ 
    SC3: Datasets to Algorithms & indirect & indirect\\
    SC4: Algorithms to Datasets & indirect & indirect\\
    SC5: Datasets to Datasets & indirect & \xmark \\
    SC6: Algorithms to Algorithms & indirect & \xmark \\ \bottomrule
    \end{tabular}
    \end{small}
    \label{tab:differences}
\end{table}

In addition to traditional item-to-user recommendations, also item-to-item recommendations can occur in data and algorithm sharing (see Figure~\ref{fig:scenarios}).
However, items, i.e., datasets and algorithms, do not directly interact with each other; a user has to run an algorithm on a given dataset.
Therefore, we rely on direct user-to-dataset and user-to-algorithm interactions to indirectly interconnect datasets and algorithms.


\textbf{SC3: Datasets to Algorithms.} 
In SC3, recommendations help to identify suitable datasets to train a given algorithm.
This scenario can occur when, e.g., algorithm providers or researchers aim to improve their algorithm via leveraging more datasets for training.
In contrast to SC1 and SC2, indirect item-to-item interaction data has to be used to generate recommendations.
Since users interact with algorithms and datasets, we can use user interactions to connect datasets and algorithms. 
Specifically, the profile and ground truth data of an algorithm consists of the datasets that users used to train the algorithm. 

\textbf{SC4: Algorithms to Datasets.} 
In SC4, recommendations help to identify suitable algorithms that can be applied to a given dataset.
This scenario can occur when dataset providers or researchers aim to find other algorithms that can be applied to their dataset, e.g., to extract a different kind of knowledge from the data.
Similar to SC3, the profile and ground truth data of a dataset consists of the algorithms that users run on the dataset. 

Finally, with SC5 and SC6, we have two additional item-to-item recommendation scenarios, but this time, the same item types are interlinked (e.g., datasets are recommended for a given dataset), which leads to a different situation with respect to the available ground truth data: 

\textbf{SC5: Datasets to Datasets.} 
In SC5, data providers can find other datasets that are utilized by the same user-community.
This is an important scenario in data economies, as this can identify datasets of competing data providers.
For SC5, we build the profile data in the same way as in case of SC4, i.e., the profile of a dataset consists of the algorithms that users run on the dataset. However, for building the ground truth data, we cannot use this idea, since we need a set of relevant datasets for a given dataset.  
To build this set, we create a collaboration network, similarly as in~\cite{kowald2019using}. This means that we create a link between two datasets if they have been used by the same user. Thus, the ground truth data of a given dataset consists of the datasets that have the largest user overlap with this dataset. 

\textbf{SC6: Algorithms to Algorithms.} 
In this scenario, algorithm providers can find other algorithms that are utilized by the same user-community.
Similar to SC5, this is an important scenario in AI-driven  economies, as this can identify algorithms of competing algorithm providers.
In case of SC6, we use the same idea as in case of SC5 to create profile and ground truth data. Hence, the profile of an algorithm consists of the datasets that users' used to train this algorithm and the ground truth data consists of the algorithms with the largest user overlap.





\section{Method} \label{s:method}

\subsection{OpenML Dataset}
Due to the lack of available datasets to evaluate recommender systems for data and algorithm sharing, we gather data from the dataset and algorithm sharing platform OpenML, and share it with the research community.

\textbf{Data crawling.} In OpenML, users can upload datasets, algorithms, or their entire machine learning pipelines, e.g., a user $u$ applied an algorithm $a$ (e.g., a classification algorithm) to a dataset $d$.
This represents user interactions between datasets and algorithms. Additionally, OpenML provides a powerful and convenient Python-based API\footnote{\url{https://docs.openml.org/Python-API/}}, which makes it an ideal platform to evaluate our recommendation scenarios.  
Our data crawling procedure is as follows:
\begin{enumerate}
    \item We use \textit{openml.datasets.list\_datasets} to fetch all datasets and \textit{openml.flows.list\_flows} to obtain all algorithms alongside their textual descriptions. Herein, we ignore case sensitivity and apply stemming.
    \item Then, we fetch triples containing the user, algorithm, and task (e.g., classification) with \textit{openml.runs.list\_runs} to obtain user interactions and  retrieve the dataset to which the user applied the algorithm by querying  \textit{openml.tasks.OpenMLTask}.
    \item Since users can apply the same algorithm to the same dataset multiple times, we cope with these repeated interactions~\cite{wang2019modeling} by merging all repetitions.
\end{enumerate}
After these three steps, our novel dataset includes 8,637,795 interactions between 544 users, 2,186 datasets, and 5,660 algorithms, as well as 2,104 datasets and 11,037 algorithms without user interactions.

However, we notice that there exist users with an extraordinarily large number of interactions.
Through close inspection, we observe that these users are used to test the platform (e.g., bots).
Thus, we remove all users, whose number of interactions exceeds the point of maximal curvature of the logarithmic-transformed  interaction-distribution~\cite{satopaaFindingKneedleHaystack2011}.
Further descriptive statistics of our OpenML dataset $\mathcal{D}$ can be found in Table~\ref{tab:datasets}.
Plus, to foster the reproducibility of our research, we provide the dataset freely via Zenodo\footnote{\url{https://doi.org/10.5281/zenodo.6517031}}. 

\textbf{Train- and test-set split.} To evaluate the performance of our recommendation methods for our recommendation scenarios, we randomly split the interactions of each target entity (i.e., user, dataset, or algorithm depending on the scenario) in our dataset $\mathcal{D}$ into 80\% profile data used for training and 20\% ground truth data used for testing.  
With this, each target entity has data for training and testing, which enables a meaningful evaluation. 
As described in Section~\ref{s:scenarios}, in SC3 and SC4, we create this profile and ground truth data using indirect item-to-item interactions (i.e., via user interactions). In SC5 and SC6, we create ground truth data via constructing a collaboration network~\cite{kowald2019using}. This means that we connect two datasets (SC5) or two algorithms (SC6)  if they were used by the same user, and put the 10 datasets or algorithms into the test set with the largest user overlap.


\begin{table}[!t]
    \centering
    \caption{Descriptive statistics of our OpenML dataset $\mathcal{D}$. Many datasets and algorithms have no interactions. Thus, in contrast to content-based recommendation methods, interaction-based recommendation methods cannot recommend these datasets and algorithms.}
    \begin{small}
    \begin{tabular}{l r}
    \toprule
        Users & 512 \\
        Algorithms & 1,307  \\
        Datasets & 573  \\
        Interactions & 10,945  \\
        Avg. Interactions / User & 21.38 \\
        Avg. Interactions / Algorithm & 8.37 \\
        Avg. Interactions / Dataset & 19.10 \\ \midrule
        Algorithms$_{w/o\ Int.}$ & 11,037  \\
        Datasets$_{w/o\ Int.}$ & 2,104  \\ \bottomrule
    \end{tabular}
    \end{small}
    \label{tab:datasets}
\end{table}

\subsection{Recommendation Methods}
\label{sec:recalg}
In the following, we present three recommendation methods~\cite{ricci2011introduction} that we evaluate in our six recommendation scenarios for data and algorithm sharing.  
Furthermore, we note that all methods generate recommendation lists of size $n=10$ and that the recommendation lists do not contain items that the target entity already knows, i.e., we filter items in the recommendation lists for which the target entity already has interactions in the profile data. 
We calculate all recommendations using the Java-based recommendation framework ScaR (Scalable Recommendation-as-a-service)~\cite{lacic2013utilizing,lacic2015scar}, and based on three built-in recommendation methods that we adapt to our data and algorithm sharing problem:

\textbf{Most Popular (MP).} 
This unpersonalized approach recommends datasets or algorithms that users interact with the most, i.e., the most popular datasets or algorithms.
This way, only a small set of available items can be recommended, even though other items may be better suited for the target entity. However, MP is always capable of recommending items, even in user cold-start settings~\cite{lacic2015tackling}. 

\textbf{Collaborative Filtering (CF).} 
CF exploits direct or indirect interaction data between users, data, and algorithms within the data- and AI-driven  economy.
For example, CF recommends a dataset $d$ to a user $u$, if similar users ($u$'s neighbors) have interacted with $d$.  
Thus, CF provides personalized recommendations generated for a target entity. 
Our CF variant is a user-based $k$-nearest neighbor approach with $k = 40$ neighbors, which is the default setting in the ScaR framework~\cite{lacic2013utilizing,lacic2015scar}. 

\textbf{Content-based Filtering (CB).} 
MP and CF rely on interaction data to generate recommendations and, therefore, are prone to popularity bias~\cite{bellogin2017statistical,abdollahpouri2019popularity}.  
As a remedy, CB generates personalized recommendations by leveraging content data, i.e., textual descriptions, to identify datasets or algorithms that are deemed to be relevant for the target entity. 
We implement CB by using TF-IDF representations~\cite{jones1972statistical,achananuparp2008evaluation} of the description text of datasets and algorithms. 
Here, we set the minimum term frequency to 1 and the minimum document frequency to 2, which are the default settings in the ScaR framework~\cite{lacic2013utilizing,lacic2015scar}.

\begin{table*}[t]
    \centering
    \caption{Our results show that CF provides the most accurate recommendations in all six recommendation scenarios. However, as $\mathrm{Cov}@10$ indicates, CF can only recommend a small fraction of the item catalog (i.e., datasets or algorithms). In contrast, CB can recommend the largest fraction of the item catalog, and provides the least popularity-biased recommendations.}
    \begin{small}
    \begin{tabular}{l r | r r r r r | r r}
        \toprule
        Recommendation Scenario & Method & $\mathrm{P}@1$ & $\mathrm{R}@10$ & $\mathrm{MRR}@10$ & $\mathrm{MAP}@10$ & $\mathrm{nDCG}@10$ & $\mathrm{Cov}@10$ & $\mathrm{RecPop}@10$ \\ \midrule
        \multirow{3}{*}{SC1 (Datasets to Users)} & MP & 0.00 & 0.22 & 0.04 & 0.04 & 0.08 & 0.01 & 593.79 \\
        & CF & \textbf{0.26} & \textbf{0.34} & \textbf{0.26} & \textbf{0.27} & \textbf{0.30} & 0.06 & 181.50 \\
        & CB & 0.05 & 0.05 & 0.03 & 0.02 & 0.04 & \textbf{0.12} & \textbf{10.25} \\ \midrule
        \multirow{3}{*}{SC2 (Algorithms to Users)} & MP & 0.03 & 0.11 & 0.05 & 0.05 & 0.07 & 0.00 & 265.75 \\
        & CF & \textbf{0.12} & \textbf{0.26} & \textbf{0.14} & \textbf{0.14} & \textbf{0.18} & 0.02 & 90.51 \\
        & CB & 0.02 & 0.06 & 0.02 & 0.03 & 0.03 & \textbf{0.03} & \textbf{9.25} \\ \midrule
        \multirow{3}{*}{SC3 (Datasets to Algorithms)} & MP & 0.00 & 0.12 & 0.02 & 0.02 & 0.04 & 0.01 & 555.20 \\
        & CF & \textbf{0.33} & \textbf{0.39} & \textbf{0.28} & \textbf{0.32} & \textbf{0.35} & 0.06 & 143.36 \\
        & CB & 0.00 & 0.13 & 0.06 & 0.06 & 0.09 & \textbf{0.14} & \textbf{7.07} \\ \midrule
        \multirow{3}{*}{SC4 (Algorithms to Datasets)} & MP & 0.01 & 0.29 & 0.12 & 0.13 & 0.18 & 0.00 & 270.62 \\
        & CF & \textbf{0.52} & \textbf{0.56} & \textbf{0.42} & \textbf{0.45} & \textbf{0.51} & 0.01 & 97.56 \\
        & CB & 0.01 & 0.03 & 0.01 & 0.01 & 0.02 & \textbf{0.03} & \textbf{12.75} \\ \midrule
        \multirow{3}{*}{SC5 (Datasets to Datasets)} & MP & 0.00 & 0.02 & 0.01 & 0.01 & 0.01 & 0.00 & 650.23 \\
        & CF & \textbf{0.17} & \textbf{0.44} & \textbf{0.17} & \textbf{0.20} & \textbf{0.28} & 0.09 & 55.74 \\
        & CB & 0.05 & 0.12 & 0.05 & 0.06 & 0.08 & \textbf{0.28} & \textbf{14.88} \\ \midrule
        \multirow{3}{*}{SC6 (Algorithms to Algorithms)} & MP & 0.01 & 0.02 & 0.01 & 0.01 & 0.01 & 0.00 & 278.32 \\
        & CF & \textbf{0.07} & \textbf{0.24} & \textbf{0.08} & \textbf{0.09} & \textbf{0.14} & 0.02 & 55.01 \\
        & CB & 0.04 & 0.12 & 0.04 & 0.05 & 0.07 & \textbf{0.04} & \textbf{7.87} \\ \midrule
    \end{tabular}
    \end{small}
    \label{tab:results}
\end{table*}
\subsection{Evaluation Criteria}
We evaluate our three recommendation methods based on two evaluation criteria: (i) accuracy, and (ii) popularity bias:

\textbf{Accuracy.} 
To evaluate recommendation accuracy, we use five widely-used metrics~\cite{parra2013recommender}: Precision $\mathrm{P}@k$, Recall $\mathrm{R}@k$, Mean Reciprocal Rank $\mathrm{MRR}@k$, Mean Absolute Precision $\mathrm{MAP}@k$, and Normalized Discounted Cumulative Gain $\mathrm{nDCG}@k$~\cite{jarvelin2002cumulated}.
Here, $\mathrm{P}@k$ is the fraction of recommended items that are relevant, 
$\mathrm{R}@k$ is the fraction of relevant items that are recommended, 
$\mathrm{MRR}@k$~\cite{radev2002evaluating} is the average reciprocal position of the relevant items in target entities' recommendation lists, 
$\mathrm{MAP}@k$ measures the quality of the ranked recommendation list by penalizing relevant items that occur later in the ranking, and 
$\mathrm{nDCG}@k$ also takes the ranking into account but is based on cumulative gain~\cite{wang2013theoretical}.

\textbf{Popularity Bias.} 
To evaluate popularity bias, we use two metrics: (i) Item Space Coverage~\cite{silveira2019good} ($\mathrm{Cov}@k$) and (ii) Average Recommendation Popularity ($\mathrm{RecPop}@k$). 
$\mathrm{Cov}@k$ is the fraction of the item catalog that is recommended to at least one target entity, and $\mathrm{RecPop}@k$ is the average popularity of the recommended items.
An item's popularity is given by the number of interactions for this item.

\section{Results}
\label{s:results}
In this section, we present the results of our experiments, in which we evaluate three recommendation methods in six recommendation scenarios along our two evaluation criteria (i) accuracy, and (ii) popularity bias.

\subsection{Accuracy} \label{s:acc}
Across our three recommendation methods Most Popular (MP), Collaborative Filtering (CF), and Content-based Filtering (CB), in Table~\ref{tab:results}, we observe that CF provides the most accurate recommendations in all our six recommendation scenarios. The most accurate recommendations across our six recommendation scenarios can be generated by CF in SC4 (Algorithms to Datasets), while the recommendations generated in SC6 (Algorithms to Algorithms) are the least accurate.
This is interesting since SC4 is a recommendation scenario, in which profile and ground truth data can only be constructed using indirect item-to-item interactions, as discussed in Section~\ref{s:scenarios}. 
However, in SC4, there exists a large item catalog (i.e., 1,307 algorithms) that CF can recommend for a few target entities (i.e., 573 datasets). 
This small dataset-to-algorithm ratio can positively impact accuracy, since selected neighbors (i.e., similar datasets) used for generating recommendations tend to be more reliable due to more co-interacted algorithms~\cite{adomavicius2012impact}.

In SC2 (Algorithms to Users), the same item catalog can be recommended to a similarly small number of target entities - in this case 512 users.
However, recommendation accuracy is substantially smaller than for SC4.
As shown in Table~\ref{tab:datasets}, users and datasets have a similar average number of interactions, but 50\% of users have more than 6 interactions, while only 28\% of datasets have more than 6 interactions.
This suggests that generating recommendations for users is more difficult than generating recommendations for datasets, possibly due to users' larger profile data.

In the case of SC6 (Algorithms to Algorithms), the recommender system needs to cope with the largest item catalog across our six recommendation scenarios.
Also, recommendations need to be generated for the very same large set of items (i.e., 1,307 algorithms can be recommended to 1,307 algorithms).
Due to this sparse interaction space, all our three recommendation methods seem to struggle with providing accurate recommendations. However, the high accuracy in SC5 (Datasets to Datasets) shows that our approach for generating ground truth data in cases where the same item type is recommended to the same item type is suitable for evaluating recommendations for data and algorithm sharing. 

\subsection{Popularity Bias} \label{s:pop}
\begin{figure}[!t]
    \centering
    \includegraphics[width=\linewidth]{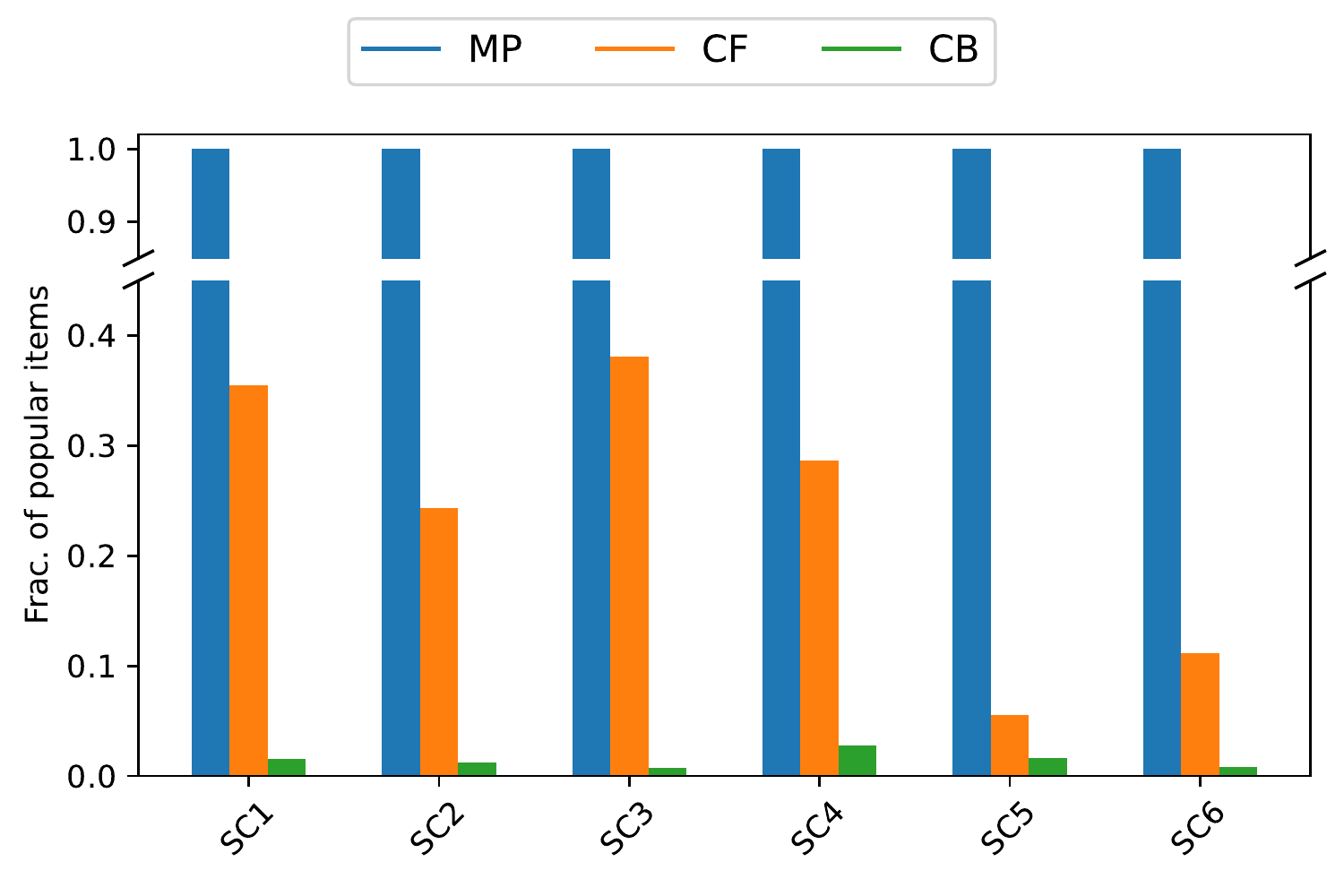}
    \caption{The fraction of popular items in a user's recommendation list. 
    MP generates the most popularity-biased recommendations, while CB recommends the least popular items.
    Also, CF tends to recommend both, popular and non-popular items.\vspace{-5mm}}
    \label{fig:popularity_bias}
\end{figure}

In our popularity bias experiments, $\mathrm{Cov}@10$ in Table~\ref{tab:results} suggests that MP can only recommend a small fraction of the item catalog, since MP always recommends the - in our case - 10 most popular items to each target entity.
This way, the exploration potential of the item catalog, which represents the datasets and algorithms in the data and AI-driven economy, is limited.
In contrast to MP, CB recommends the largest fraction of the item catalog in all six recommendation scenarios and, therefore, allows exploring a larger part of the data- and AI-driven economy.
Plus, CB generates the least popularity-biased recommendations, since $\mathrm{RecPop}@10$ exhibits a smaller value than in case of MP and CF.

In Figure~\ref{fig:popularity_bias}, we discuss popularity bias in more detail and investigate the fraction of recommendations for popular items (i.e., the 10 most popular items a target entity has not rated yet).
Similar to our results in Table~\ref{tab:results}, we can observe that in all six recommendation scenarios, MP provides the most popularity-biased recommendations, while CB tends to recommend items with low popularity. 
In contrast to the interaction-based recommendation methods MP and CF, the content-based recommendation method CB can recommend items without interactions, i.e., cold-start items. 

In general, CB mostly recommends non-popular items and MP recommends only popular items. CF tends to recommend both, popular and non-popular items and thus, provides more popularity-balanced recommendations.
Our finding that CF provides more accurate and popularity-balanced recommendations than MP and CB is in line with recent research that shows that accurate recommendations should also take non-popular items into account in addition to popular items~\cite{abdollahpouri2019unfairness,kowald2020unfairness,naghiaei2022unfairness}.

\section{Conclusion}
In this work, we evaluate the applicability of recommender systems for supporting data and algorithm sharing.
Therefore, we create a novel dataset based on the OpenML dataset and algorithm sharing platform, to enable an offline evaluation of three standard recommendation methods in six recommendation scenarios.
Plus, we discuss our results along two criteria: recommendation accuracy and popularity bias.
We find that Collaborative Filtering can generate more accurate dataset and algorithm recommendations than Most Popular and Content-based Filtering.
Moreover, Content-based Filtering exhibits popularity bias to the smallest extent and can recommend many of the datasets and algorithms that are ignored by Most Popular and Collaborative Filtering.
Overall, our work discusses how recommender systems can be applied within data- and AI-driven economies to support data and algorithm sharing.

\textbf{Limitations and future work.}
We recognize one main limitation of this work: we do not investigate the aspect of monetization of data and algorithm sharing in data- and AI-driven economies.
Instead, our work focuses on data and algorithm sharing itself, and how recommender systems can support the interconnection of users, data, and algorithms.
However, we acknowledge that, e.g., a recommended dataset might be relevant for a given user, but could exceed the user's financial possibilities. 
Thus, developing recommender systems that are aware of financial constraints remains an interesting avenue for future research.
Moreover, in this work, we focus on three broad families of recommender systems, i.e., popularity-, collaboration-, and content-based approaches.
However, our future work will also incorporate more specialized approaches as, e.g., deep learning or matrix factorization.
Also, we will acknowledge that data and algorithm providers may have privacy-related, legal, ethical, or economical concerns when making resources available through recommendations.
Thus, we will work on how these concerns can be respected in a recommender system, e.g., by incorporating privacy-preserving technologies. 

\textbf{Acknowledgements.} This research was supported by the H2020 project TRUSTS (GA: 871481), and the ``DDIA'' and ``DDAI'' FFG COMET Modules. 

\bibliographystyle{ACM-Reference-Format}
\bibliography{bibliography}

\end{document}